\documentclass[letterpaper]{article} % DO NOT CHANGE THIS
\usepackage{aaai2026}  % DO NOT CHANGE THIS
\usepackage{times}  % DO NOT CHANGE THIS
\usepackage{helvet}  % DO NOT CHANGE THIS
\usepackage{courier}  % DO NOT CHANGE THIS
\usepackage[hyphens]{url}  % DO NOT CHANGE THIS
\usepackage{graphicx} % DO NOT CHANGE THIS
\usepackage{natbib}  % DO NOT CHANGE THIS AND DO NOT ADD ANY OPTIONS TO IT
\usepackage{caption} % DO NOT CHANGE THIS AND DO NOT ADD ANY OPTIONS TO IT
\usepackage{algorithm}
\usepackage{algorithmic}
\usepackage{amsmath}
\usepackage{amssymb}  
\usepackage{booktabs}
\usepackage{multirow}
\usepackage{graphicx} % for \resizebox
\usepackage{float}
\usepackage{subcaption}

\usepackage{newfloat}
\usepackage{listings}
\DeclareCaptionStyle{ruled}{labelfont=normalfont,labelsep=colon,strut=off} % DO NOT CHANGE THIS
\floatstyle{ruled}
\newfloat{listing}{tb}{lst}{}
\floatname{listing}{Listing}
\title{Unbiased Rectification for Sequential Recommender Systems Under Fake Orders}
\author{
    Qiyu Qin, Yichen Li, Haozhao Wang, Cheng Wang,
    Rui Zhang, 
    Ruixuan Li\thanks{Ruixuan Li is the corresponding author.}
}
\affiliations{
    School of Computer Science and Technology, Huazhong University of Science and Technology, Wuhan, China\\
    \{qiyu\_qin, ycli0204\}@hust.edu.cn, rayteam@yeah.net$\langle$www.ruizhang.info$\rangle$ \\  
}

\usepackage{bibentry}
\begin{document}

\maketitle

\begin{abstract}
Fake orders pose increasing threats to sequential recommender systems by misleading recommendation results through artificially manipulated interactions, including click farming, context-irrelevant substitutions, and sequential perturbations. 
Unlike injecting carefully designed fake users to influence recommendation performance, fake orders embedded within genuine user sequences aim to disrupt user preferences and mislead recommendation results, thereby manipulating exposure rates of specific items to gain competitive advantages. 
To protect users' authentic interest preferences and eliminate misleading information, this paper aims to perform precise and efficient rectification on compromised sequential recommender systems while avoiding the enormous computational and time costs of retraining existing models. Specifically, we identify that fake orders are not absolutely harmful—in certain cases, partial fake orders can even have a data augmentation effect.
Based on this insight, we propose \textit{Dual-view Identification} and \textit{Targeted Rectification (DITaR)}, which primarily identifies harmful samples to achieve unbiased rectification of the system. The core idea of this method is to obtain differentiated representations from collaborative and semantic views for precise detection, and then filters detected suspicious fake orders to select truly harmful ones for targeted rectification with gradient ascent. 
This ensures that useful information in fake orders is not removed while preventing bias residue. Moreover, it maintains the original data volume and sequence structure, thus protecting system performance and trustworthiness to achieve optimal unbiased rectification.
Extensive experiments on three datasets demonstrate that DITaR achieves superior performance compared to state-of-the-art methods in terms of recommendation quality, computational efficiency, and system robustness. 
\end{abstract}

\begin{links}
    \link{Code}{https://github.com/QinWHang/DITaR}
\end{links}

\section{Introduction}

\begin{figure}[t]
    \centering
    \includegraphics[width=\linewidth]{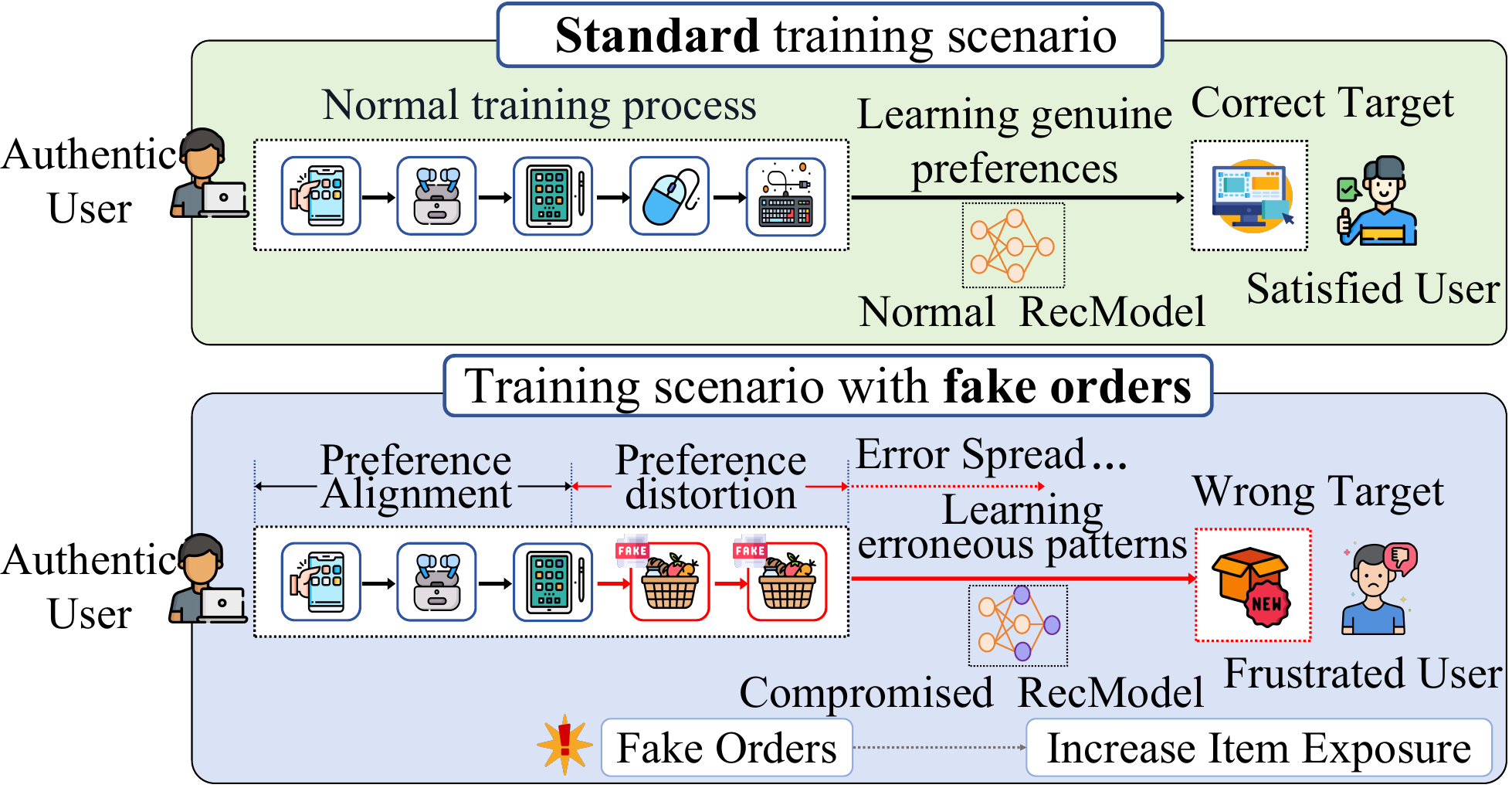}
    \caption{Comparison between standard sequential recommendation process and sequential recommendation process under fake orders. Fake orders alter recommendation results, leading to degraded user experience.}
    \label{fake orders}
\end{figure}

Sequential recommender systems have become a core component of modern recommendation systems by modeling users' historical interaction sequences to predict their future preferences~\cite{intro1,intro2,zhang_seqmlp}. Unlike traditional collaborative filtering methods, sequential recommender systems capture the dynamic changes in user interests and temporal dependencies between items, playing important roles in e-commerce~\cite{intro3}, social media~\cite{intro4}, and streaming platforms~\cite{intro5}.
\par
With the growing commercial value of sequential recommender systems, artificially manipulated interaction behaviors generated for specific commercial purposes deserve attention. These behaviors are strategically embedded in genuine user interaction sequences, including click farming, context-irrelevant semantic substitutions, and sequential perturbations, which we term \textit{fake orders}. These covert fake orders steer users towards merchant-preferred items by disrupting the collaborative filtering patterns and semantic associations essential for accurate sequential modeling.
This manipulation distorts authentic user preferences, introduces systematic bias, generates erroneous recommendations, and severely damages user trust in the system. 
Unlike previous adversarial approaches that inject carefully designed fake users to degrade overall system performance~\cite{intro_attack1, intro_attack2}, fake orders exploit system trust in seemingly legitimate historical users to manipulate specific recommendation results. 
% However, fake orders, by their artificial nature, struggle to maintain consistency across both collaborative and semantic dimensions simultaneously. Manipulating one view to appear plausible inevitably leaves detectable anomalies in the other, enabling cross-view identification. 
However, fake orders, by their artificial nature, struggle to maintain consistency across both collaborative and semantic dimensions simultaneously. This inherent limitation creates detectable cross-view discrepancies that enable their identification.
Figure.\ref{fake orders} illustrates how fake orders infiltrate genuine user sequences, alter recommendation results, and degrade user experience.
\par
Data rectification has emerged as a promising strategy to eliminate harmful data samples while enabling models to focus on meaningful information. In recommender systems, most rectification methods adopt user-wise or item-wise strategies~\cite{intro_wise}. To improve precision, clustering-based approaches shard datasets and use attention-weighted aggregation for targeted retraining~\cite{exp_setup_RecEraser}. For efficiency, influence function methods estimate data impact through gradient approximation~\cite{intro_inf_function, feng_unlearning}. Model-specific strategies also exist, including hierarchical deletion operators for GNNs~\cite{intro_graph,graph_unlearning} and reverse learning that modifies objectives to forget manipulated data~\cite{intro_rrl}.

\par
While existing rectification methods achieve promising results by removing noisy data through collaborative filtering approaches, ideal rectification should not simply remove all noisy data. Instead, it should gain deeper insights into the real impact of noisy data, removing truly harmful data while preserving beneficial information. Furthermore, when confronting the specific challenges posed by fake orders, existing methods face significant limitations. \textbf{C1:} Most existing methods are designed for collaborative filtering models and struggle with sequential systems where temporal dependencies and evolving user preferences create complex interdependencies that complicate efficient rectification. \textbf{C2:} How to achieve sample-wise rectification for fake orders in sequential models. Since fake orders can be injected at low cost without altering data volume, simple removal disrupts data integrity, requiring precise impact elimination while preserving surrounding valid interactions. \textbf{C3:} Building on the need for precise rectification, the challenge is how to avoid uniform fake order treatment, quantify their actual impact, and preserve beneficial information while removing harmful samples to achieve efficient unbiased rectification.
\par
To address these challenges, we propose Dual-view Identification and Targeted Rectification (DITaR), an unbiased rectification framework for sequential recommender systems under fake orders. DITaR operates through two complementary stages: The identification phase derives differentiated representations from collaborative and semantic view and detects suspicious fake orders by analyzing cross-view representation inconsistencies and intrinsic attribute anomalies. The rectification phase employs influence function estimation to assess the actual impact of detected fake orders, filtering out truly harmful samples and applying targeted rectification with gradient ascent to achieve optimal unbiased performance.
\par
Extensive experiments on three datasets demonstrate that our proposed DITaR significantly improves recommendation performance and computational efficiency compared to state-of-the-art methods while ensuring system robustness. Our contributions are as follows:
\begin{itemize}
    \item We are the first to focus on the novel and covert scenario of fake orders embedded within genuine user sequences, which manipulate recommendation outcomes and undermine user trust, posing critical challenges for integrity.
    \item We propose a dual-view framework that exploits semantic-collaborative representation gaps for fake order identification, coupled with influence-guided filtering and gradient ascent for targeted rectification, achieving unbiased and efficient sample-level rectification.
    \item We conduct comprehensive experiments demonstrating that DITaR significantly outperforms state-of-the-art baselines across multiple evaluation metrics.
\end{itemize}

%You can remove the copyright notice and ensure that your names aren't shown by including \texttt{submission} option when loading the \texttt{aaai2026} package:

%\begin{quote}\begin{scriptsize}\begin{verbatim}
%\documentclass[letterpaper]{article}
%\usepackage[submission]{aaai2026}
%\end{verbatim}\end{scriptsize}\end{quote}

%The remainder of this document are the original camera-
%ready instructions. Any contradiction of the above points
%ought to be ignored while preparing anonymous ssubmis-
%sions.

\section{Related Works}

\paragraph{\textbf{Sequential Recommendation.}} Sequential recommendation systems aim to capture users' dynamic interests by modeling temporal dependencies and preference evolution within their interaction histories~\cite{rw_sq_aim, rw_sq_dy1, yc_ijcai}. 
Recent attention-based models have brought significant advances: SASRec~\cite{rw_sq_sas} 
leverages transformer self-attention for long-term dependencies, while BERT4Rec~\cite{rw_sq_bert} employs bidirectional encoders for comprehensive interest modeling. To achieve more refined representations, recent work integrates multimodal features~\cite{rw_sq_mm,zhang_m3oe,yc_www} and leverages large language models to transform item attributes into high-quality semantic embeddings~\cite{rw_sq_llm1, rw_sq_llm2,zhang_llmseq}.

\paragraph{\textbf{User History Manipulation.}}
While sequential data's multi-layered complexity in collaborative patterns, semantic associations, and temporal dependencies enables precise modeling, it also creates vulnerabilities to manipulation. Recent studies have demonstrated that genuine user interaction histories can be manipulated through documented mechanisms: web injection techniques tamper with in-transit webpage content to force unintended user interactions~\cite{rw_att_web}, while substitution-based approaches strategically replace vulnerable sequence elements with targeted items~\cite{rw_att_defend}.
Building on these feasibility demonstrations, this paper addresses three specific manipulation types embedded within genuine sequences, collectively termed fake orders.

\paragraph{\textbf{Rectification Methods.}} Data rectification aims to eliminate harmful samples' negative impact without prohibitive retraining costs, crucial for maintaining system performance and robustness~\cite{rw_rm_aim1, rw_rm_aim2}.
This technique has seen wide application across domains, including noisy label correction in computer vision~\cite{rw_rm_cv, yc_icml}, managing large language model lifecycles~\cite{rw_rm_llm1, rw_rm_llm2,yc_com}.
Influence functions~\cite{rw_rm_inf1} quantify individual samples' contributions to model performance~\cite{rw_rm_inf2, intro_inf_function}. Other approaches include gradient-based methods like TracIN~\cite{rw_rm_tracin}, Shapley-based methods~\cite{rw_rm_shapley1, rw_rm_shapley2}, and clustering~\cite{rw_rm_cluster} or graph-based methods~\cite{rw_rm_graph,zhang_graph} that identify anomalous samples.
In this paper, we focus on rectification for sequential recommendation, identifying truly harmful items and applying targeted rectification.

\begin{figure*}
    \centering
    \includegraphics[width=0.95\linewidth]{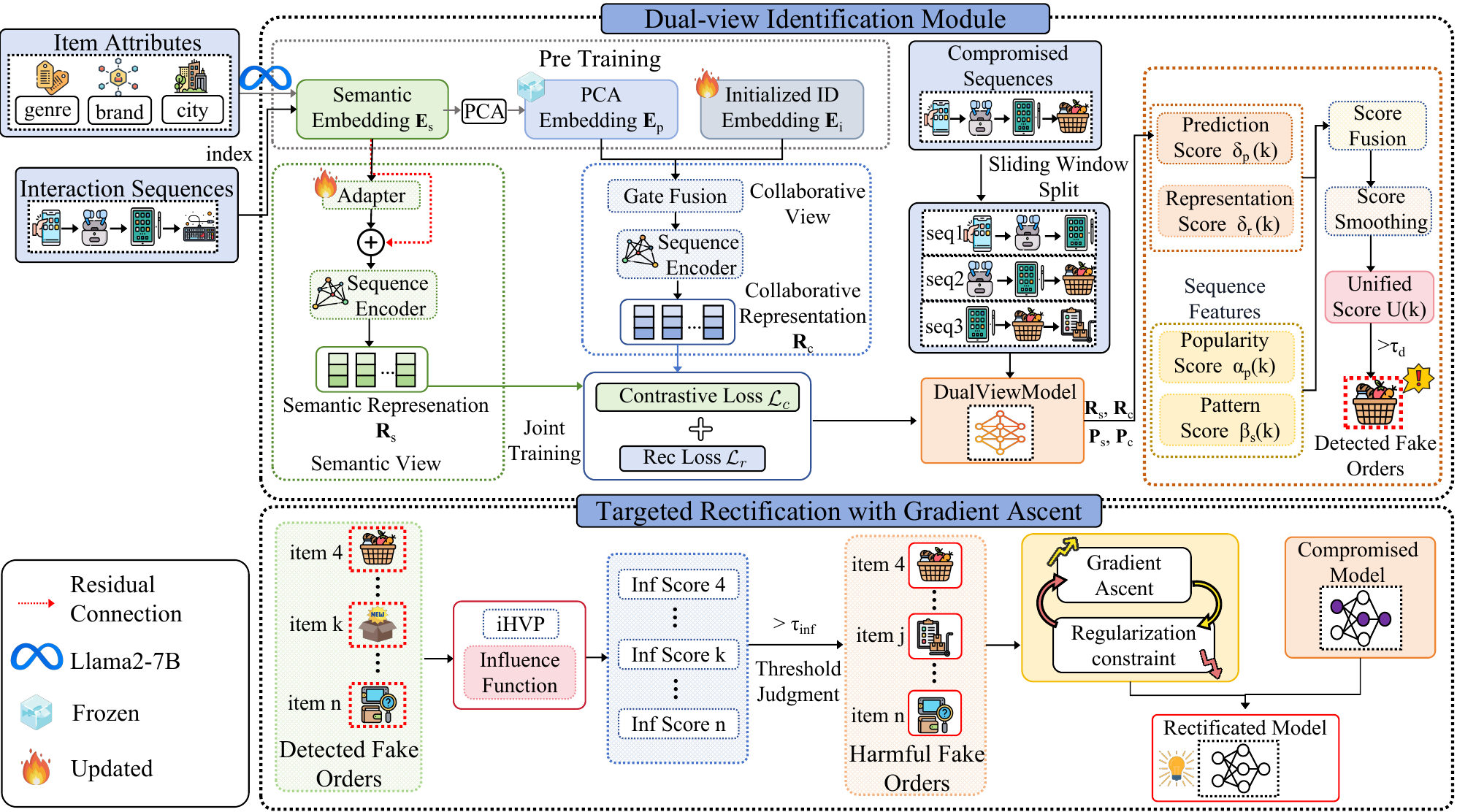}
    \caption{The framework of DITaR. DITaR first constructs collaborative and semantic view models to derive dual-view representations, which are combined with sequence features for fake order detection. Subsequently, influence function assesses the true impact of detected samples on model parameters, enabling selective gradient ascent rectification on harmful fake orders with appropriate regularization to achieve the final rectified model.}
    \label{framework}
\end{figure*}

\section{Methodology}
This section formalizes the definitions of sequential recommendation and Hessian Matrix, then proposes the Dual-view Identification and Targeted Rectification (DITaR) framework. The framework consists of two core components: Dual-view Identification (DI) Module and Targeted Rectification (TaR) with Gradient Ascent. DI identifies fake orders by combining differentiated information from collaborative and semantic views. TaR filters items based on influence function and performs precise rectification on harmful fake orders. Figure.\ref{framework} illustrates the overall DITaR framework.

\subsection{Problem Formulation}
\noindent\textbf{Sequential Recommendation.} Given a user's action sequence $\mathcal{U}=\left\{u_1, u_2, \ldots, u_m\right\}$ and item sequence $ \mathcal{I}=\left\{i_1, i_2, \ldots, i_n\right\}$, sequential recommendation systems aim to predict the next item that user $u$ is likely to interact with, based on the user's historical interaction sequence $S_u=\left[i_1^u, i_2^u, \ldots, i_{\left|S_u\right|}^u\right]$. The model parameters $\theta$ are trained by minimizing the loss function
\begin{equation}
	\theta^* = \arg\min_{\theta} \sum_{u \in \mathcal{U}} \sum_{t=1}^{|S_u|-1} \mathcal{L}(f(S_u^{1:t}; \theta), i_{t+1}^u).
\end{equation}

\noindent\textbf{Hessian Matrix.} Influence function requires computing the product of the inverse Hessian matrix and vector v: $H_\theta^{-1} v$. Given model $\theta$, $H_{\theta}$ represents the Hessian matrix~\cite{mt_inf_funct}, i.e., the second-order derivative matrix of the loss function with respect to model parameters. It is defined as: 
\begin{equation}
    H_\theta=\nabla_\theta^2 \sum_j \mathcal{L}\left(i_j, \theta\right).
\end{equation}
For recommendation systems with large numbers of parameters, directly computing $H_\theta^{-1}$ is infeasible. With $p$ as parameter dimension and $J$ as recursion depth, we approximate $H_\theta^{-1} v$ through implicit Hessian-vector products (IHVP)~\cite{rw_rm_inf1}, reducing computational complexity from $O\left(p^2\right)$ to $O(p \cdot J)$. 
\par
\subsection{ Dual-view Identification Module}
We consider complementary information from collaborative and semantic views. The collaborative view reflects statistical patterns of user-item interactions, while the semantic view explores intrinsic associations between items.Fake orders, constrained by their artificial generation mechanisms, struggle to maintain consistency across both dimensions simultaneously, creating 
detectable cross-view discrepancies that distinguish them from genuine interactions where both views naturally align. Through deep mining of dual-view information and analysis by a unified detection framework, DITaR accurately identifies different types of fake orders.
\par
We learn separated collaborative filtering patterns and semantic association features through representation learning and contrastive decoupling. For the semantic view, item attributes and descriptive texts are organized into complete prompts and fed into pre-trained LLaMA2-7B~\cite{mt_llama2} to extract semantic embedding $\mathbf{E}_{s} \in \mathbb{R}^{d}$, where $d$ is LLaMA2's embedding dimension. To align features between language model and recommendation space, we employ a semantic adapter for feature transformation while preserving original semantics through residual connection. 
For the collaborative view, we apply Principal Component Analysis (PCA) to $\mathbf{E}_{s}$ to obtain dimension-reduced embedding $\mathbf{E}_{p}$, which is then fused with learnable ID embedding $\mathbf{E}_{i}$ through an adaptive gating mechanism. This mechanism balances collaborative patterns with item-specific personalization.
Finally, to preserve distinct characteristics and prevent information leakage, separate sequential encoders $f_{s}$ and $f_{c}$ process each view independently to generate disentangled representations $\mathbf{R}_{s}$ and $\mathbf{R}_{c}$.
\begin{equation}
       \begin{gathered}
\mathbf{R}_s=f_s\left[\operatorname{Adapter}\left(\mathbf{E}_s\right)+\lambda_1 \cdot\left(w_1 \mathbf{E}_s\right)\right], \\[5pt]
    \mathbf{R}_c=f_c\left[\mathbf{G} \odot\left(w_2 \mathbf{E}_p+b_1\right)+(1-\mathbf{G}) \odot \mathbf{E}_i\right].
    \end{gathered}
\end{equation}
where $\text{Adapter}(\cdot)$ is a two-layer nonlinear network, $\lambda_1$ is the residual coefficient, $\mathbf{G} \in \mathbb{R}^{d_h}$ is the adaptive gating vector, $\odot$ denotes element-wise multiplication, and $w_1, w_2, b_1$ are learnable parameters.
We optimize the model through a joint training objective that combines a recommendation loss with a contrastive loss. Specifically, we employ the InfoNCE loss as the contrastive learning objective to explicitly enforce the independence of the two views. The contrastive loss $\mathcal{L}_c$ is defined as:
\begin{equation}
    \begin{aligned}
    \mathcal{L}_c & =-\frac{1}{2 B} \sum_{i=1}^B\left[l_{\mathrm{s}}(i)+l_{\mathrm{c}}(i)\right] ,\\
    \text { where } l_{\mathrm{s}}(i) & =\log \frac{\exp \left(\hat{\mathbf{R}}_s(i) \cdot \hat{\mathbf{R}}_c(i) / \tau\right)}{\sum_{j=1}^B \exp \left(\hat{\mathbf{R}}_s(i) \cdot \hat{\mathbf{R}}_c(i) / \tau\right)}.
    \end{aligned}
\end{equation}
where $\hat{\mathbf{R}}=\mathbf{R}/|\mathbf{R}|_2$ represents $\ell_2$ normalization, $\tau$ is the temperature coefficient, and $B$ denotes batch size.
%With $\alpha$ controlling the weights of the two views, considering the recommendation loss $\mathcal{L}_r$, the final joint optimization objective $\mathcal{L}_t$ is:
The final joint optimization objective $\mathcal{L}_t$ combines the recommendation loss $\mathcal{L}_r$ and contrastive loss $\mathcal{L}_c$:
\begin{equation}
  \mathcal{L}_t=\alpha \mathcal{L}_r^s+(1-\alpha) \mathcal{L}_r^c+\lambda_2 \mathcal{L}_c.
\end{equation}
where $\alpha$ controls the weight between the two views and $\lambda_2$ controls the importance of contrastive learning for view independence. We set $\alpha = 0.5$ to balance dual-view contributions equally and $\lambda_2 = 0.1$ to provide moderate contrastive regularization.
The joint optimization strategy ensures that dual views maintain feature independence while collaboratively completing recommendation tasks, thereby providing a more discriminative representational basis for detecting manipulated interactions. 
\par
Based on the trained dual-view model, we construct a unified analysis framework to identify fake orders. The core principle leverages cross-view information asymmetry: genuine interactions should maintain inherent consistency across collaborative and semantic views, while fake orders, due to their artificial manipulation nature, generate detectable behavioral inconsistency signals across different dimensions.
Our detection strategy operates on the insight that fake orders manifest themselves through multiple complementary anomaly patterns. To capture these diverse signals comprehensively, we decompose the detection process into cross-view consistency analysis and intrinsic behavioral pattern analysis. For each interaction $i_k$ in the sequence, we extract its dual-view representations ${\mathbf{R}_s(k), \mathbf{R}_c(k)}$ and prediction distributions ${P_s(k), P_c(k)}$ from the trained model.
The first component focuses on cross-view inconsistencies that arise when fake orders disrupt the natural alignment between collaborative and semantic signals. We measure representation disagreement $\delta_r(k)$ using cosine similarity and prediction divergence $\delta_p(k)$ through Jensen-Shannon Divergence~\cite{mt_jsd}, as these metrics effectively capture the fundamental conflicts introduced by artificial manipulation.
The second component analyzes intrinsic behavioral anomalies by examining popularity patterns and contextual disruptions. We quantify popularity anomaly $\alpha_p(k)$ through statistical z-score deviation and contextual disruption $\beta_s(k)$ via local sequential pattern consistency analysis. Importantly, these intrinsic features help distinguish fake orders from natural behavioral variations such as exploratory clicks: while both may show interest shifts, genuine variations are grounded in historical co-occurrence patterns, whereas fake orders lack such foundations and thus exhibit anomalies across both cross-view and intrinsic dimensions simultaneously.
These complementary dimensions are integrated through an adaptive weighting mechanism that learns the relative importance of each signal for fake order detection. Specifically, the multi-dimensional anomaly feature vector $\mathbf{f}_k = [\delta_p(k), \delta_r(k), \alpha_p(k), \beta_s(k)]$ captures these four complementary signals. The unified anomaly score is computed as:
\begin{equation}
    u(k)=\mathbf{w}^T \odot \mathcal{N}\left(\mathbf{f}_k\right).
\end{equation}
where $\mathbf{w}$ is the learned weight vector, and $\mathcal{N}(\cdot)$ denotes score normalization.
To enhance robustness against local noise and ensure stable detection performance, we apply temporal smoothing regularization to obtain the final decision score $U(k)$ for each position. An adaptive threshold $\tau_d$ is automatically tuned on a validation set to optimize detection performance. This framework achieves precise and robust identification of fake orders through systematic multi-dimensional signal analysis, providing a high-quality candidate set for subsequent influence-based rectification.
\subsection{Targeted Rectification with Gradient Ascent} Given a compromised recommendation model trained on data containing fake orders, our goal is to rectify it efficiently without incurring the substantial cost of retraining, altering data volume, or causing system instability. To this end, we first leverage influence function to identify truly harmful fake orders, then perform a targeted gradient ascent to precisely neutralize their negative effects. This approach ensures system performance and robustness, achieving our goal of unbiased rectification.
We begin with the insight that not all identified fake orders are detrimental to the model's performance. Therefore, for each suspicious interaction $i_k$ in the detected fake orders set $I_d$, we quantify its true impact on the model's performance on a clean validation set using influence function~\cite{mt_inf}. 
The influence of an item $i_k$, denoted as $\mathrm{Inf}(i_k)$, can be expressed as: 
\begin{equation}
    \mathrm{Inf}\left(i_k\right)=-g_{v}^T H_\theta^{-1} \nabla_\theta \mathcal{L}\left(i_k, \theta\right).
\end{equation}
where $\nabla_\theta \mathcal{L}\left(i_k, \theta\right)$ is the loss gradient of the suspicious sample, $H_\theta$ is the Hessian matrix, and $g_v$ is the average gradient over the clean validation set.
% Positive values indicate the fake order is harmful to system performance, while negative values indicate benefit. The harmful interaction set $I_h$ is:
The influence function approximates how removing $i_k$ would affect validation performance. Positive values indicate removing $i_k$ decreases validation loss, thus $i_k$ is harmful; negative values suggest beneficial augmentation effects.
\begin{equation}
    {I}_h=\left\{i_k \in {I}_{\text {d}}: \operatorname{Inf}\left(i_k\right)>\tau_{\text {inf }}\right\}.
\end{equation}
Typically, $\tau_{\text{inf}}$ is set to 0, ensuring that only interactions confirmed to be harmful are targeted for rectification, thereby fulfilling the objective of unbiased rectification. 
To precisely and efficiently remove the negative impact of the identified harmful fake orders $I_h$, we perform a targeted gradient ascent step. Let $\theta$ be the current model parameters. We compute the intermediate parameters $\theta_m$ as follows:
\begin{equation}
    \theta_m=\theta_t+\eta_1 \sum_{i_k \in {I}_h } \nabla_\theta \mathcal{L}\left(i_k, \theta_t\right).
\end{equation}
where $\eta_1$ is the rectification learning rate. Meanwhile, we apply regularization constraint by performing one-step gradient descent using clean dataset ${D}_c$ after each update. With regularization learning rate $\eta_2 \ll \eta_1$, we ensure the model maintains modeling capability for normal recommendation tasks while rectifying harmful information.
\begin{equation}
   \theta_{t+1}=\theta_m-\eta_2 \cdot \frac{1}{\left|{D}_c\right|} \sum_{j \in {D}_c} \nabla_\theta \mathcal{L}\left(j, \theta_m\right).
\end{equation}
Through alternating optimization strategy and monitoring validation performance to prevent overfitting, we finally output the rectified model $\theta_r$, achieving balance between system security and recommendation performance to accomplish unbiased rectification.

\section{Experiment}
In this section, we evaluate our proposed method using three datasets and various baselines. Our analysis assesses DITaR's rectification performance in terms of recommendation effectiveness, efficiency, and robustness. We then investigate the true impact of fake orders and perform ablation studies on DITaR's core detection and rectification modules to demonstrate the overall effectiveness of our approach from multiple perspectives.

\subsection{Experiment Setup}
\paragraph{\textbf{Datasets.}} We conduct experiments on three datasets: MovieLens-1M (ML-1M)~\cite{exp_setup_ml}, Amazon-Beauty, and Yelp2018~\cite{exp_setup_yelp}. Following~\cite{rw_sq_llm2}, we apply standard preprocessing to filter out users and items with insufficient interactions.
We design three fake order scenarios corresponding to the manipulation types introduced earlier: (1) Repetitive orders simulate click farming by repeating a selected item for k subsequent positions, disrupting collaborative filtering patterns; (2) Semantic orders replace target items with semantically irrelevant items, breaking semantic consistency; (3) Sequential orders alter the interaction order of two non-adjacent items within a specified window, corrupting temporal dependencies.
The number of affected items is controlled by user ratio (proportion of affected users) and intensity (proportion of fake orders per affected user). Fake orders replace genuine interactions without changing sequence length, and we use mutually exclusive allocation to ensure only one type of fake order is applied per position, guaranteeing detection reliability by avoiding interference between different types. 
Following the dataset configuration of SASRec~\cite{rw_sq_sas}, we adopt a leave-one-out evaluation protocol where the last item in each user's interaction sequence constitutes the test set, while the second-to-last item forms the validation set. 
For evaluation, we randomly sample negative items that users have not interacted with, which are then paired with the ground truth positive item to compute ranking-based metrics.
Dataset statistics and fake order configurations are presented in Table \ref{tab:dataset_stats}.

\begin{table}[t]
    \centering
    \begin{minipage}{0.47\textwidth}
    \renewcommand\arraystretch{1.4}
    \resizebox{\linewidth}{!}{
    \begin{tabular}{lcccc}
        \toprule
        Dataset & \#Users & \#Items & \#Interactions & \#Fake Orders \\
        \midrule
        ML-1M           & 6,040     & 3,416    & 999,611 & 140,258 (14.03\%) \\
        Amazon-Beauty   & 22,363    & 12,101   & 198,502 & 23,603 (11.89\%) \\
        Yelp2018        & 213,170 & 94,304 & 3,277,932 & 43,165 (13.18\%) \\
        \bottomrule
    \end{tabular}}
    \caption{Dataset statistics and Fake Orders settings with user ratio=0.3, intensity =0.3.}
    \label{tab:dataset_stats}
    \end{minipage}
\end{table}

\paragraph{\textbf{Baselines \& Recommendation Models.}} 
We select four rectification methods as baselines: Retrain, SISA~\cite{exp_setup_SISA}, RecEraser~\cite{exp_setup_RecEraser}, and UltraRE~\cite{exp_setup_UltraRE}. Retrain removes fake orders and retrains from scratch; SISA utilizes sharding aggregation by dividing data into shards and averaging predictions from all sub-models to obtain aggregated scores; RecEraser introduces clustering for diversified shard partitioning with attention-weighted aggregation; UltraRE extends RecEraser using Optimal Balanced Clustering (OBC) and simplifies aggregation logic. We set the number of shards to 10 following original implementations. These 
baselines, primarily designed for collaborative filtering, are adapted to sequential recommendation tasks. We implement these baselines and our method on classic sequential recommendation models including SASRec~\cite{rw_sq_sas}, GRU4Rec~\cite{exp_setup_gru}, and BERT4Rec~\cite{rw_sq_bert}.
\paragraph{\textbf{Configurations.}} 
Fake orders are configured in two groups: (1) 30\% user ratio with 30\% intensity, and (2) 60\% user ratio with 60\% intensity, where each setting includes all three fake order types. For each baseline model, we configure according to original paper recommendations. For evaluation metrics, we use Hit Rate (HR)@k and NDCG@k with default k values of 10 and 20. To evaluate rectification efficiency and avoid differences caused by different training parameter designs, we use convergence epochs for assessment. For experimental robustness, we report average results from two independent runs with random seeds \{42, 43\}.

\begin{table*}[ht]
    \centering
    \renewcommand\arraystretch{1.3}
    \resizebox{\linewidth}{!}{
        \begin{tabular}{llcccccccccccc}
            \toprule
            \multicolumn{2}{c}{}
            & \multicolumn{4}{c}{SASRec} & \multicolumn{4}{c}{GRU4Rec} & \multicolumn{4}{c}{Bert4Rec} \\
                \cmidrule(r){3-6} \cmidrule(r){7-10} \cmidrule(r){11-14}
                Dataset & Method 
                  & N@10 & H@10 & N@20 & H@20 
                  & N@10 & H@10 & N@20 & H@20 
                  & N@10 & H@10 & N@20 & H@20 \\
                \midrule
                \multirow{6}{*}{ML-1M}
                  & Original & 0.2449 & 0.4329 & 0.2908 & 0.6154 & 0.2440 & 0.4288 & 0.2876 & 0.6025 & 0.2790 & 0.5111 & 0.3306 & 0.7152 \\
                  & Retrain & 0.2406 & 0.4311 & 0.2829 & 0.5988 & 0.2365 & 0.4207 & 0.2806 & 0.5960 & 0.2666 & 0.4975 & 0.3154 & 0.6902 \\
                  & SISA & 0.1089 & 0.2185 & 0.1402 & 0.3443 & 0.0818 & 0.1523 & 0.1110 & 0.2682 & 0.2411 & 0.4404 & 0.2871 & 0.6242 \\
                  & RecEraser & 0.2035 & 0.3813 & 0.2483 & 0.5598 & 0.1803 & 0.3356 & 0.2235 & 0.5075 & 0.2355 & 0.4296 & 0.2786 & 0.6013 \\
                  & UltraRE & 0.2120 & 0.3732 & 0.2526 & 0.5348 & 0.1104 & 0.1907 & 0.1381 & 0.3012 & 0.2440 & 0.4434 & 0.2895 & 0.6238 \\
                  & \textbf{DITaR}(ours) & \textbf{0.2470} & \textbf{0.4386} & \textbf{0.2893} & \textbf{0.6076} & \textbf{0.2448} & \textbf{0.4369} & \textbf{0.2854} & \textbf{0.5978} & \textbf{0.2763} & \textbf{0.5036} & \textbf{0.3267} & \textbf{0.7025} \\
                \midrule
                \multirow{6}{*}{Amazon-Beauty} 
                  & Original & 0.2036 & 0.3356 & 0.2344 & 0.4583 & 0.1809 & 0.3159 & 0.2140 & 0.4470 & 0.2029 & 0.3601 & 0.2361 & 0.4918 \\
                  & Retrain & 0.1971 & 0.3320 & 0.2268 & 0.4502 & 0.1735 & 0.3087 & 0.2054 & 0.4352 & 0.1912 & 0.3462 & 0.2243 & 0.4777 \\
                  & SISA & 0.1023 & 0.1999 & 0.1318 & 0.3173 & 0.1040 & 0.1989 & 0.1338 & 0.3168 & 0.1232 & 0.2179 & 0.1573 & 0.3527 \\
                 & RecEraser & 0.1817 & 0.3192 & 0.2140 & 0.4475 & 0.1630 & 0.2890 & 0.1937 & 0.4114 & 0.1779 & 0.3124 & 0.2082 & 0.4327 \\
                 & UltraRE & 0.1768 & 0.3113 & 0.2092 & 0.4396 & 0.1724 & 0.3050 & 0.2044 & 0.4318 & 0.1593 & 0.2969 & 0.1948 & 0.4380 \\
                  & \textbf{DITaR}(ours) & \textbf{0.2032} & \textbf{0.3364} & \textbf{0.2331} & \textbf{0.4555} & \textbf{0.1725} & \textbf{0.3070} & \textbf{0.2060} & \textbf{0.4400} & \textbf{0.1971} & \textbf{0.3523} & \textbf{0.2300} & \textbf{0.3527} \\
                \midrule
                \multirow{6}{*}{Yelp2018} 
                  & Original & 0.3161 & 0.5155 & 0.3485 & 0.6433 & 0.3174 & 0.5187 & 0.3509 & 0.6510 & 0.4034 & 0.6453 & 0.4389 & 0.7851 \\
                  & Retrain & 0.3069 & 0.4932 & 0.3373 & 0.6130 & 0.2734 & 0.4547 & 0.3067 & 0.5870 & 0.3869 & 0.6225 & 0.4231 & 0.7650 \\
                  & SISA & 0.2009 & 0.3609 & 0.2349 & 0.4948 & 0.2070 & 0.3449 & 0.2369 & 0.4632 & 0.2968 & 0.4971 & 0.3333 & 0.6417 \\
                  & RecEraser & 0.2753 & 0.4634 & 0.3098 & 0.6000 & 0.2327 & 0.3974 & 0.2652 & 0.5259 & 0.3296 & 0.5415 & 0.3670 & 0.6891 \\
                  & UltraRE & 0.2795 & 0.4553 & 0.3105 & 0.5783 & 0.2881 & 0.4746 & 0.3224 & 0.6102 & 0.3515 & 0.5673 & 0.3861 & 0.7039 \\
                  & \textbf{DITaR}(ours) & \textbf{0.3128} & \textbf{0.5003} & \textbf{0.3434} & \textbf{0.6215} & \textbf{0.3106} & \textbf{0.5072} & \textbf{0.3441} & \textbf{0.6397} & \textbf{0.3924} & \textbf{0.6309} & \textbf{0.4282} & \textbf{0.7720} \\
                \bottomrule
        \end{tabular}}
    \caption{Performance comparison of various methods for the top-k recommendation task. The best results are bold.}
    \label{tab:Performance Overall}
\end{table*}

\subsection{Performance Overall}
\paragraph{\textbf{Rectification Results.}} We evaluate the effectiveness, robustness, and efficiency of DITaR. Table\ref{tab:Performance Overall} demonstrates the rectification performance of different methods across three datasets. DITaR consistently outperforms baseline methods and achieves performance comparable to retraining, occasionally even surpassing the original clean data performance. This indicates that DITaR successfully balances data preservation by removing harmful components while retaining beneficial information for high-quality rectification.
The performance gap between retraining and the original method reveals the critical impact of data loss in recommender systems. As fake order injection frequency and intensity increase, this degradation becomes more severe, highlighting the limitations of simple data removal approaches. 
Existing sharding-based methods attempt to balance rectification efficiency and accuracy through partitioning and model aggregation. However, these approaches face inherent limitations. Sharding inevitably disrupts collaborative relationships between users, while aggregation introduces approximation errors that cannot fully recover the original collaborative signals. Moreover, these methods were primarily designed for traditional collaborative filtering models and struggle with the complex temporal dependencies in sequential data. Their uniform treatment of all detected samples overlooks the heterogeneous nature of fake orders, failing to distinguish between harmful and potentially beneficial instances.

\par
In contrast, DITaR maintains performance levels close to the original clean data with minimal fluctuation, demonstrating strong resistance to fake order intrusions while preserving both system robustness and recommendation quality. The framework's consistent performance across different datasets and model architectures validates its effectiveness and demonstrates stability compared to baseline methods.

\begin{table}[t]
    \centering
    \begin{minipage}{0.47\textwidth}
    \centering
    \resizebox{\linewidth}{!}{
    \begin{tabular}{llccc}
        \toprule
        \multicolumn{2}{c}{} & \multicolumn{3}{c}{Model} \\
        \cmidrule(r){3-5}
        Dataset & Method & SASRec & GRU4Rec & Bert4Rec \\
        \midrule
        \multirow{5}{*}{ML-1M}
            & Retrain & 140 & 100 & 130 \\
            & SISA & 60 & 80 & 67 \\
            & RecEraser & 52 & 55 & 50 \\
            & UltraRE & 35 & 38 & 43 \\
            & \textbf{DITaR}(ours) & 5 & 5 & 5 \\
        \midrule
        \multirow{5}{*}{Amazon-Beauty}
            & Retrain & 175 & 110 & 125 \\
            & SISA & 86 & 78 & 66 \\
            & RecEraser & 43 & 50 & 51 \\
            & UltraRE & 38 & 45 & 36 \\
            & \textbf{DITaR}(ours) & 5 & 5 & 5 \\
        \midrule
        \multirow{5}{*}{Yelp2018}
            & Retrain & 145 & 135 & 180 \\
            & SISA & 119 & 81 & 120 \\
            & RecEraser & 55 & 57 & 60 \\
            & UltraRE & 43 & 55 & 50 \\
            & \textbf{DITaR}(ours) & 5 & 5 & 5 \\
        \bottomrule
    \end{tabular}}
    \caption{Convergence performance of different rectification methods under fake orders (user ratio = 0.3, intensity = 0.3).}
    \label{tab:Convergence}
    \end{minipage}
\end{table}

\par
Table \ref{tab:Convergence} compares the computational efficiency of different methods using convergence epochs as a fair evaluation metric. For sharding-based methods, we report the rounded average convergence epochs of sub-models across different shards with evaluation performed every 5 epochs. DITaR demonstrates significantly fewer convergence epochs compared to baseline methods. While sharding-based approaches may achieve faster training on individual small models due to reduced data volume, their overall efficiency decreases as the number of shards increases, creating a trade-off between training efficiency and accuracy. Moreover, more uniform sharding strategies lead to smoother convergence across sub-models, with similar convergence epochs among different shards. DITaR benefits from its targeted rectification paradigm, requiring only gradient ascent fine-tuning on pre-trained models rather than training from scratch, thus achieving superior computational efficiency.

\begin{figure}[t]
    \centering
    \begin{subfigure}[t]{0.48\linewidth}
        \centering
        \includegraphics[width=\linewidth]{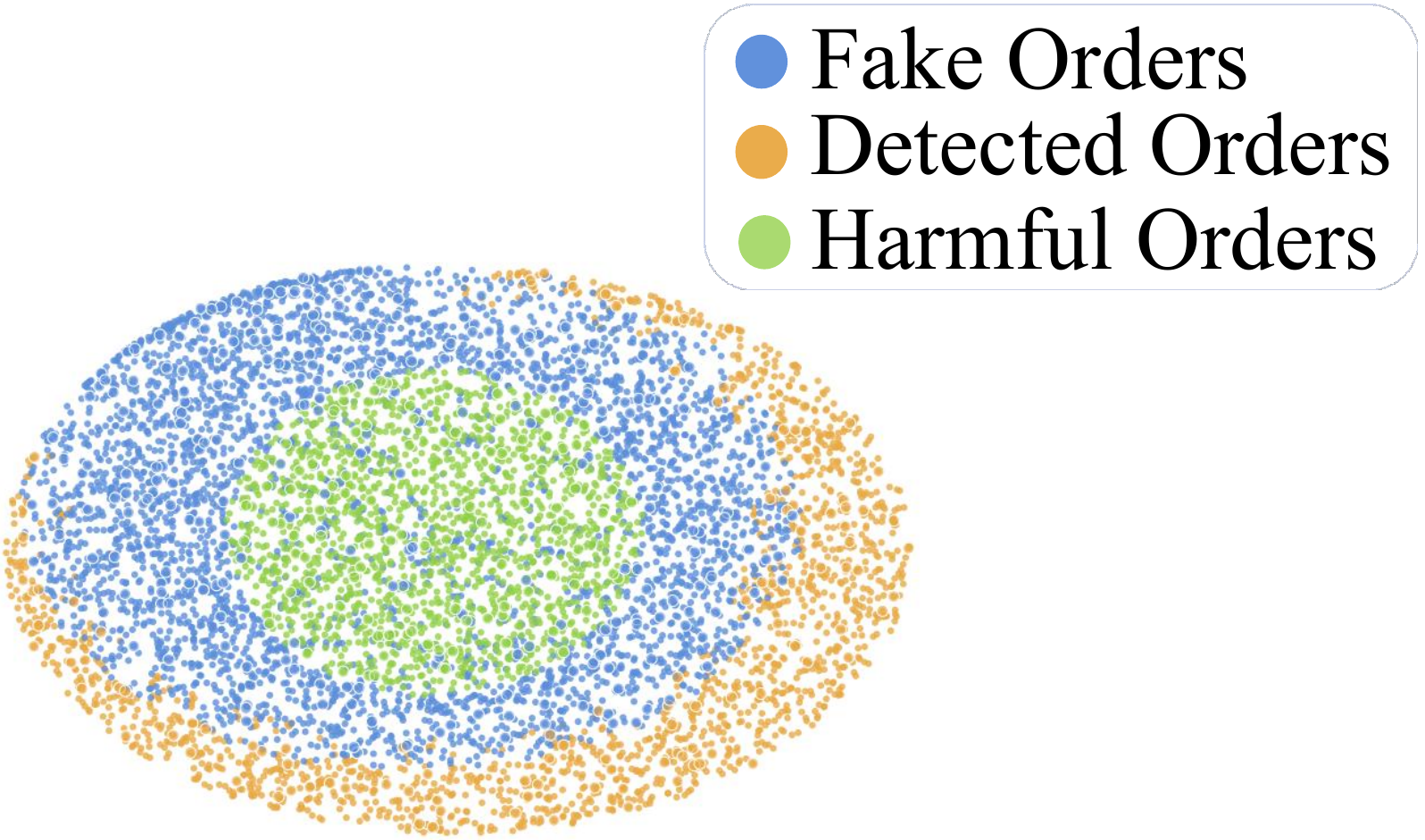}
        \caption{Detection with SASRec on ML-1M.}
        \label{sas_ml_detection}
    \end{subfigure}
    \hfill
    \begin{subfigure}[t]{0.48\linewidth}
        \centering
        \includegraphics[width=\linewidth]{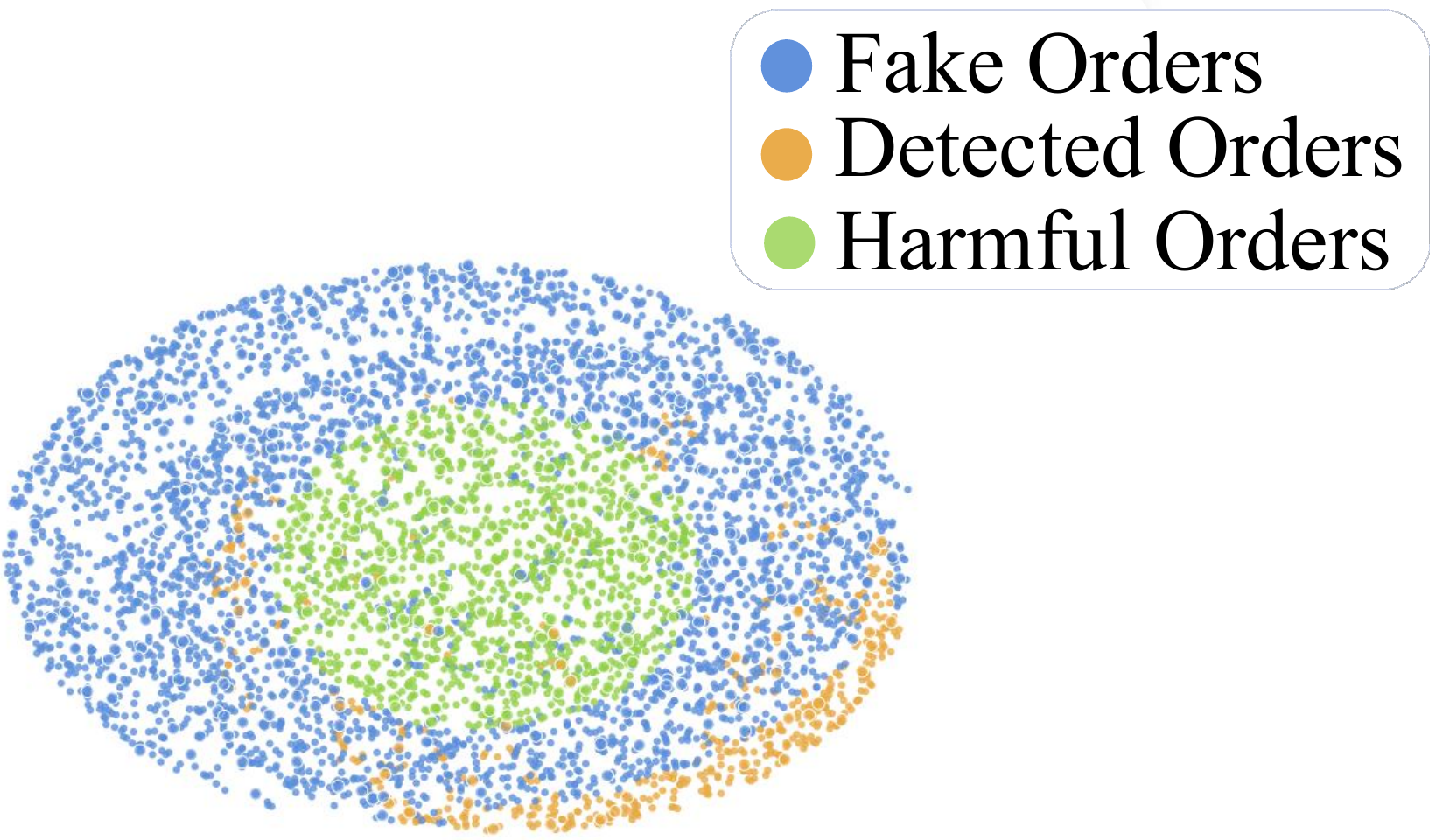}
        \caption{Detection with Bert4Rec on Amazon-Beauty.}
        \label{bert_beauty_detection}
    \end{subfigure}
    \caption{Fake order detection performance of two sequential recommendation models across different datasets.}
    \label{fig:detection}
\end{figure}

\paragraph{\textbf{Detection Effectiveness.}} Figure.\ref{fig:detection} illustrates the performance of our detection module. Through dual-view modeling and the unified detection framework, our method gains deep insights into the information disruptions caused by fake orders from sequences' collaborative patterns and semantic coherence. The results demonstrate that DITaR achieves high precision while maintaining strong recall, successfully capturing the majority of fake orders despite occasional false positives or missed detections. This provides a reliable candidate set for subsequent targeted rectification, establishing a solid foundation for precise fake order removal.

\begin{figure}[htbp]
    \centering
    \begin{subfigure}[t]{0.49\linewidth}
        \centering
        \includegraphics[width=\linewidth]{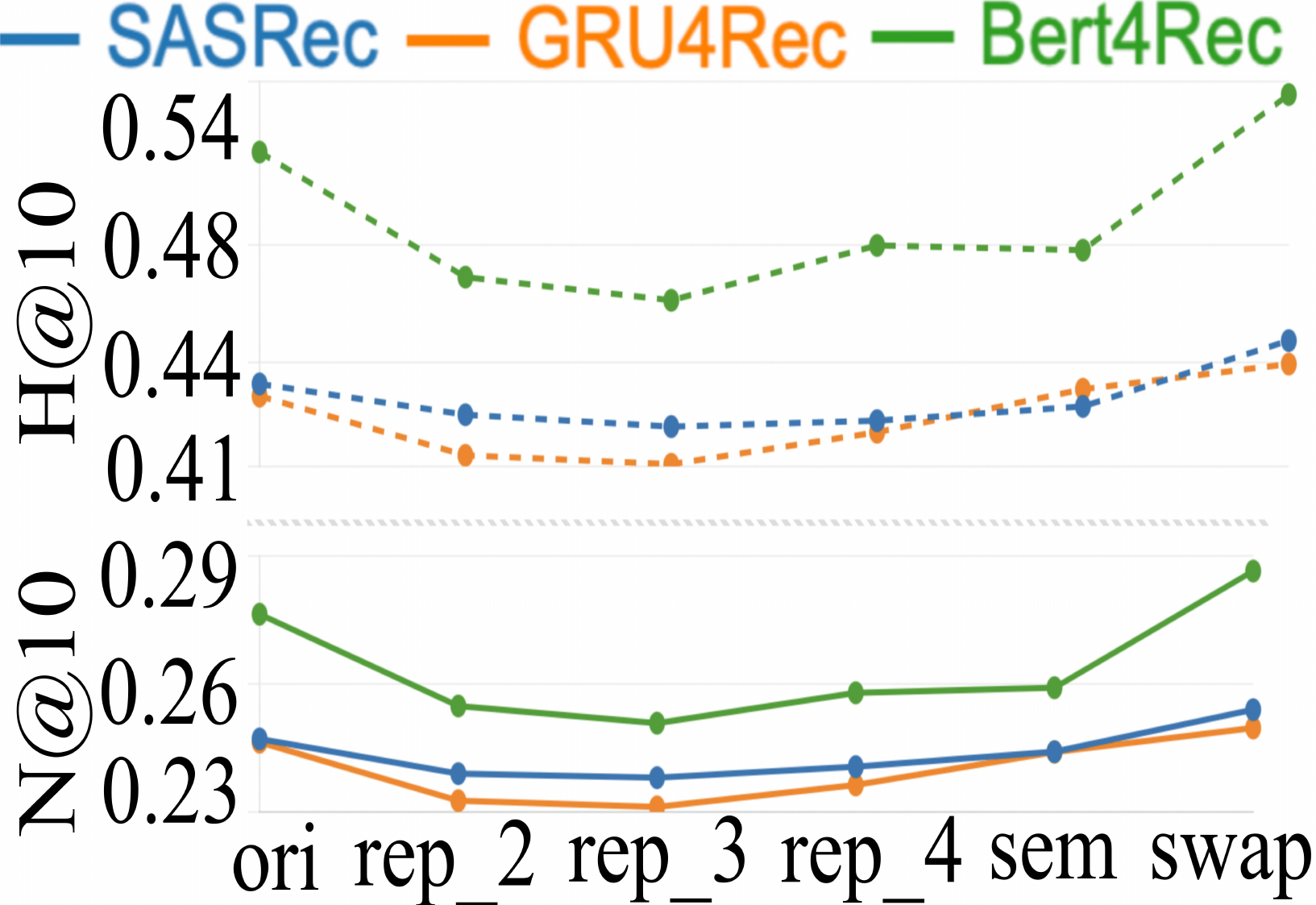}
        \caption{Fake Orders Effects on ML-1M with user ratio=0.3, intensity=0.3.}
        \label{sas_ml_fake_order_effect}
    \end{subfigure}
    \hfill
    \begin{subfigure}[t]{0.49\linewidth}
        \centering
        \includegraphics[width=\linewidth]{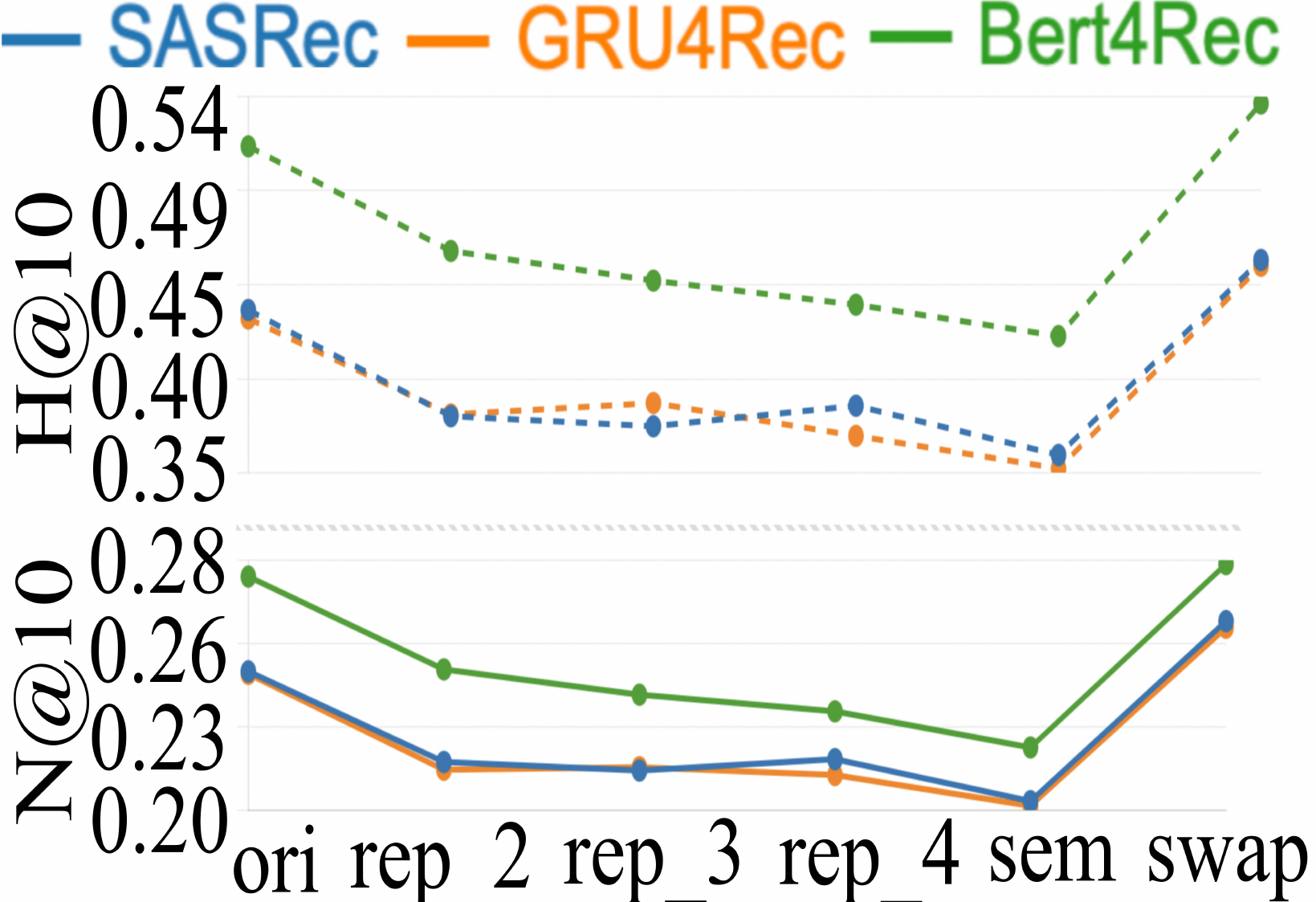}
        \caption{Fake Orders Effects on ML-1M with user ratio=0.6, intensity=0.6.}
        \label{bert_beauty_fake_order_effect}
    \end{subfigure}
    \caption{Impact of fake order types on recommendation performance (ori: original data; rep\_X: repetitive orders with length X; sem: semantic orders; swap: sequential orders).}
    \label{fig:effcts of fake orders}
\end{figure}

\paragraph{\textbf{Real Impact of Fake Orders.}} To validate our core insight that "not all fake orders are harmful," we analyze the individual impact of different fake order types. Figure.\ref{fig:effcts of fake orders} reveals distinct effects across the three scenarios. Repetitive orders demonstrate performance fluctuations as repeat length increases, with moderate repetition potentially reinforcing certain interaction patterns. Semantic orders consistently degrade system performance due to their disruption of semantic coherence, showing pronounced negative effects. Most notably, sequential order swapping demonstrates a beneficial effect on system performance. This counterintuitive finding can be attributed to the fact that swapping non-adjacent items introduces controlled temporal noise that acts as implicit data augmentation, helping the model learn more robust sequential patterns and reducing overfitting to rigid temporal dependencies. This regularization effect enhances the model's generalization capability, particularly for handling naturally occurring temporal variations in user behavior. The impact magnitude correlates positively with both user ratio and intensity parameters, indicating that widespread and frequent fake order injections pose greater threats to system integrity and warrant heightened attention.

\begin{table}[h]
    \centering
    \resizebox{\linewidth}{!}{
    \begin{tabular}{lcccccc}
        \toprule
        \multicolumn{1}{c}{} 
        & \multicolumn{2}{c}{SASRec} 
        & \multicolumn{2}{c}{GRU4Rec} 
        & \multicolumn{2}{c}{Bert4Rec} \\
        \cmidrule(r){2-3} \cmidrule(r){4-5} \cmidrule(r){6-7}
        Method & N@10 & H@10 & N@10 & H@10 & N@10 & H@10 \\
        \midrule
        \textbf{DITaR} & 0.2032 & 0.3364 & 0.1725 & 0.3070 & 0.1971 & 0.3523 \\
        -w/o Semantic View & 0.1865 & 0.3108 & 0.1566 & 0.2788 & 0.1802 & 0.3226 \\
        -w/o Collaborative View & 0.1829 & 0.3061 & 0.1534 & 0.2736 & 0.1753 & 0.3153 \\
        -w/o Influence Function & 0.1941 & 0.3240 & 0.1641 & 0.2924 & 0.1862 & 0.3338 \\
        \bottomrule
    \end{tabular}}
    \caption{Ablation study results on Amazon-Beauty dataset across SASRec, GRU4Rec, and BERT4Rec models.}
    \label{tab:ablation}
\end{table}

\paragraph{\textbf{Ablation Study.}} As shown in Table \ref{tab:ablation}, we evaluate the complementary effects of dual views and influence function filtering capabilities. Removing either view leads to decreased final rectification effectiveness, with collaborative view removal showing more significant impact. This demonstrates that collaborative filtering patterns and temporal dependencies remain the core of sequential modeling, while semantic information serves as crucial complementary features that enhance sequential representation learning. Additionally, to ensure high recall value, our detection module uses relatively lenient thresholds, detecting more candidate items than actual fake orders. Without influence function filtering, directly applying gradient ascent to all detected candidates would inadvertently harm beneficial or neutral items, leading to performance degradation. This highlights the indispensable role of influence function estimation in distinguishing harmful from beneficial fake orders, thereby achieving truly unbiased rectification. Nevertheless, dual-view detection without influence filtering still outperforms single-view approaches, further validating the superior detection capability of our dual-view framework.

\section{Conclusion}
In this paper, we propose DITaR, an efficient unbiased rectification framework for sequential recommender systems under fake orders. Our approach exploits dual-view modeling to identify fake orders through cross-view discrepancies, then leverages influence function to selectively rectify genuinely harmful samples while preserving beneficial information. This enables precise sample-level rectification without retraining or altering data structures. Comprehensive experiments demonstrate that DITaR significantly outperforms the state-of-the-art 
methods across multiple evaluation metrics.

\section*{Acknowledgments}
This work is supported by the National Key Research and Development Program of China under grant 2024YFC3307900; the National Natural Science Foundation of China under grants 62376103, 62302184 and 62436003; Major Science and Technology Project of Hubei Province under grant 2025BAB011 and 2024BAA008; and Hubei Science and Technology Talent Service Project under grant 2024DJC078.

\bibliography{aaai2026}

\end{document}